\documentclass[%
 aip,
 amsmath,amssymb,
 reprint,%
]{revtex4-1}

\usepackage{graphicx}
\usepackage{dcolumn}
\usepackage{bm}
\usepackage{array}

\usepackage[utf8]{inputenc}
\usepackage[T1]{fontenc}
\usepackage{mathptmx}
\usepackage{tabularx} 
\usepackage{color,soul}

\begin{document}

\preprint{AIP/123-QED}

\title[]{Quantifying the destructuring of a thixotropic colloidal suspension using falling ball viscometry\\}

\author{Rajkumar Biswas}
\altaffiliation[]{rajkumar@rri.res.in}
\author{Debasish Saha}
\altaffiliation[]{Debasish.Saha@uni-duesseldorf.de}
\author{Ranjini Bandyopadhyay}
\altaffiliation[]{ranjini@rri.res.in}
\affiliation{Raman Research Institute, Bangalore - 560080, India
}%

\date{\today}

\begin{abstract}
The settling dynamics of falling spheres inside a Laponite suspension is studied. Laponite is a colloidal synthetic clay that shows physical aging in aqueous suspension due to the spontaneous evolution of inter-particle electrostatic interactions. In our experiments, millimeter-sized steel balls are dropped in aqueous Laponite suspensions of different ages (i.e., time elapsed since sample preparation). The motion of the falling balls are captured using a high-speed camera and the velocities of their centroids are estimated from the images. Interestingly, we observe that balls of larger diameters fail to achieve terminal velocity over the entire duration of the experiment. We propose a mathematical model that accounts for rapid structural changes (expected to be induced by the falling ball) in Laponite suspensions whose aging time scales are much slower than the time of fall of the ball. For a range of ball sizes and Laponite suspension ages, our model correctly predicts the time-dependence of the ball velocity. Furthermore, fits to our model allow us to estimate the rates of destructuring of the thixotropic suspensions due to the passage of the falling ball.
\\

\end{abstract}

\maketitle
\section{Introduction}

The settling dynamics of objects in a non-Newtonian aging suspension leads to very interesting modifications of Stokesian flows. It is well known that for an object falling through a Newtonian medium under gravity, a terminal velocity, $V_{ter}$, arises due to a balance between the surface drag on the object and its buoyant weight. For low Reynolds numbers, the flow around the surface of the object remains laminar. In this regime, the drag force is linearly proportional to the velocity. The relation between terminal velocity and drag coefficient of an object falling through a viscous medium at low Reynolds numbers $(Re<1)$ was first determined by Stokes in 1851 \cite{stokes1851}. The differential equation for a falling ball inside a Newtonian fluid can be written as, $m\dot{V} = -  \kappa V + F_b$, where $\dot{V}$ = time derivative of the instantaneous velocity $V$ of the ball, $\kappa$ = drag coefficient, $\eta$ = viscosity of the fluid, $D$ = diameter of the ball, $m$ = mass of the ball, $F_b$ = buoyant force. Solving this equation, we can write  the expression for velocity, $V$: 
\begin{equation}
V = V_{ter} + (V_0 - V_{ter})\exp(- t/\tau)   \label{one}
\end{equation}
where $V_0$ = initial velocity and $V_{ter} = (\rho_s - \rho) g D^2/18 \eta$, $\tau = m/\kappa$, where $\kappa$ = $3\pi\eta D$, $\rho_s$ = density of the ball, $\rho$ = density of the fluid, $g$ = acceleration due to gravity. For high Reynolds numbers $(10-500)$, the flow does not remain laminar around the ball even though it attains a steady state. Eqn.1 can be used for a range of Reynolds numbers by including the appropriate correction factor for drag\cite{CoulsonRichardson}. The dependence of the drag coefficient on the Reynolds number has been investigated extensively for Newtonian fluids\cite{brown2003, goossens2019, khan1987, little1976}. For intermediate Reynolds number, an empirical form of correction factor was proposed by Schiller and Naumann\cite{CoulsonRichardson, goossens2019}. It has also been reported that fluid motion around a ball remains symmetric in Newtonian fluids, while the symmetry is broken when the ball falls through an aging viscoelastic fluid with a finite yield stress\cite{gueslin2006, Freund}.

Novel flow properties such as shear thinning, shear thickening, non-zero normal stresses and yield stresses emerge in viscoelastic materials such as polymers, gels and suspensions \cite{Guvendiren2012,Bonn2010,Fong1996}. The bulk rheological response of a non-Newtonian complex fluid can be correlated with its structural details at microscopic length scales. Bio-materials like human blood\cite{Giannokostas},  collagen\cite{Clark}, silk fibre\cite{Liu} exhibit thixotropic behaviour. These complex bio-materials are used in disease detection and treatment, cell carrier therapy, bio-printing etc. The quality of common everyday materials such as toothpaste, cosmetics, paint, mayonnaise, clays etc., can also be improved by understanding their non-Newtonian flow \cite{Chhabra}. The recovery of oil from soil, which consists of different clay compounds, sand, organic matter, oil and water in varying proportions, can be enhanced with the knowledge of the flow of particles in non-Newtonian fluids \cite{Zhuang2019}.

Recently, flow patterns around objects falling through Xanthan gum suspensions have been studied\cite{Mrokowska2019}. Falling ball viscometry is a simple but elegant technique to study non-Newtonian fluids. In this method, the motion of a ball falling through a medium is tracked to obtain  information about the properties of the latter. Such experiments can effectively probe the complex flow properties of fluids whose structures continually evolve with time. In a worm-like micellar solution, the motion of a falling ball can lead to instabilities \cite{jayaraman2003, kumar2012}, with larger balls seen to oscillate in the direction of their fall due to shear induced structural deformation of the underlying medium during their passage. Oscillatory motion was also observed for rising bubbles in a worm-like micellar solution \cite{Handzy2013}. The velocity of a spherical ball in Bentonite clay suspensions was found to increase continuously over a period of time \cite{Briscoe1992}. For non-Newtonian fluids in general, observations like shear dependent viscosity, fore-aft asymmetry, negative wake and change in yielding behaviour with aging time make the motion of the falling object more complicated \cite{gueslin2006}. 

In this work, the motion of a ball falling through aging Laponite suspensions is studied. Laponite is a synthetic clay comprising disk-like particles of thickness $1$ nm and diameter $25-30$ nm \citep{Kroon1998}. In dry form, the particles exist in 1-D stacks called tactoids that start disintegrating when dissolved in water. For jammed aqueous suspensions of Laponite, complete break-up of tactoids into single particles cannot occur due to strong inter-particle electrostatic repulsions \cite{Ali2013}. The viscosity of a Laponite suspension of concentration between $2-3.5$\% $w/v$ is observed to spontaneously increase due to the evolution of the internal structures in a physical aging process. Several studies have reported that continuous variations in inter-particle interactions occur due to Na$^+$ dissociation from Laponite surfaces in aqueous suspension \cite{Tawari2001, Joshi2007}. This process proceeds simultaneously with the break-up of Laponite stacks or tactoids and gives rise to a continuous evolution of the microscopic structure of the suspension. Laponite suspensions exhibit a physical aging phenomenon wherein the suspension viscosity spontaneously increases with time due to the underlying structural buildup\cite{Bonn}. Microscopic interactions between clay particles are sensitive to clay concentration, pH of the medium and temperature \cite{Saha2015, Jatav2014}. The concentration-dependent phase diagram of aqueous Laponite suspensions has been studied extensively in the literature. These studies reveal that Laponite suspensions can exist in sol, gel, glass and even a nematic phase\cite{ruzika2011, Cummins2007}. Rearrangements between clay particles, resulting from inter-particle screened electrostatic interactions, lead to the observed sequence of phases. 

The rheological behaviour of aging Laponite suspensions has a strong dependence on the preparation protocol. It can vary noticeably between a freshly prepared sample and one that has been rejuvenated (where aging has been reinitialized by shearing a spontaneously evolving suspension). A partial irreversibility in aging behaviour upon rejuvenation is also observed\cite{Shahin2010} and understood by considering that the structural build-up in Laponite suspensions for long aging times cannot be destroyed even by applying very large shear deformations. A dichotomic aging behaviour, indicating different aging dynamics in freshly prepared and rejuvenated Laponite suspensions, was observed using X-ray photon correlation spectroscopy and dynamic light scattering\cite{Angelini2013}. It has been reported that a ball dropped in a rejuvenated Laponite suspension gets stuck inside the fluid\cite{tabuteau2007} or oscillates  during its fall\cite{Fazilati2017}. The critical stress value that has to be exerted by the ball to exhibit these contrasting behaviours depends on the ball size and density, and the  aging time, $t_w$, defined as the time since preparation of the Laponite suspension \cite{moller2006}. It is observed that the medium in the immediate neighborhood of the surface of the ball fluidizes due to the shear-induced viscous thinning that arises from the disruption of the micro-structure due to the passage of the ball. Regions far from the ball are less fluid-like in comparison\cite{Putz2008}. Most falling ball experiments performed in Laponite suspensions have used rejuvenated samples\cite{tabuteau2007, Fazilati2017}. Since falling ball viscometry has not been performed in spontaneously aging Laponite suspensions to the best of our knowledge, we have conducted a systematic study to probe the evolution in the underlying microscopic structures of freshly prepared Laponite suspensions.

In this article, balls of different sizes are dropped into freshly prepared Laponite suspensions having different initial states of structure that depend on the suspension age quantified by the waiting time, $t_w$, since sample preparation. The motion of the ball falling in Laponite suspensions is captured using a high speed camera and is observed to differ qualitatively from the trajectory of a ball falling through a Newtonian fluid. We note that while a ball falling through a Newtonian liquid of comparable viscosities reaches a steady state within the experimental time window, the velocity of a ball dropped in the Laponite suspension continues to increase, with the increase being more dramatic for the larger balls used in this study. Representative data for a 3 mm ball dropped in a Newtonian glycerol-water mixture and a non-Newtonian Laponite suspension of comparable viscosity are plotted in Fig.S1 in Supplementary Material. Interestingly, the larger balls used in this study, when dropped in the Laponite suspensions of ages 1-3 hrs explored here, do not reach a steady state over the entire duration of the experiment. We modify an existing rheological model \cite{moore1959} for time-dependent changes in the structure of an aging non-Newtonian fluid to explain our experimental observations. We show here that when the time of fall of the ball is much faster than the aging time of the medium, fits of our experimental ball-drop data to the model can be used to successfully extract the destructuring time scales of aging Laponite suspensions.

\section{Experimental Methods}

\subsection{Sample preparation}
All experiments reported in this work were performed using Laponite XLG. Laponite, which is in powdered form and hygroscopic in nature, was first heated in a hot air oven for $16-20$ hr at $120^\circ $C. Colloidal suspensions of Laponite of concentration $2.8$\% $w/v$ were prepared by adding $19.6$ gm dried Laponite powder to $700$ ml Milli-Q water. The concentration of Laponite is expressed as the weight of Laponite powder in grams that was added to $100$ ml Milli-Q water. Laponite suspensions were vigorously stirred in a magnetic stirrer for $1$ hr to disperse the Laponite powder in the aqueous medium. After $1$ hr, the suspension became optically clear. It was then filtered in a closed container through a milli-pore membrane sheet (pore size = $0.45$ $\mu$m) with the help of a vacuum pump to break up any remaining particle clusters.

\begin{figure}[ht]
\includegraphics[scale=.14]{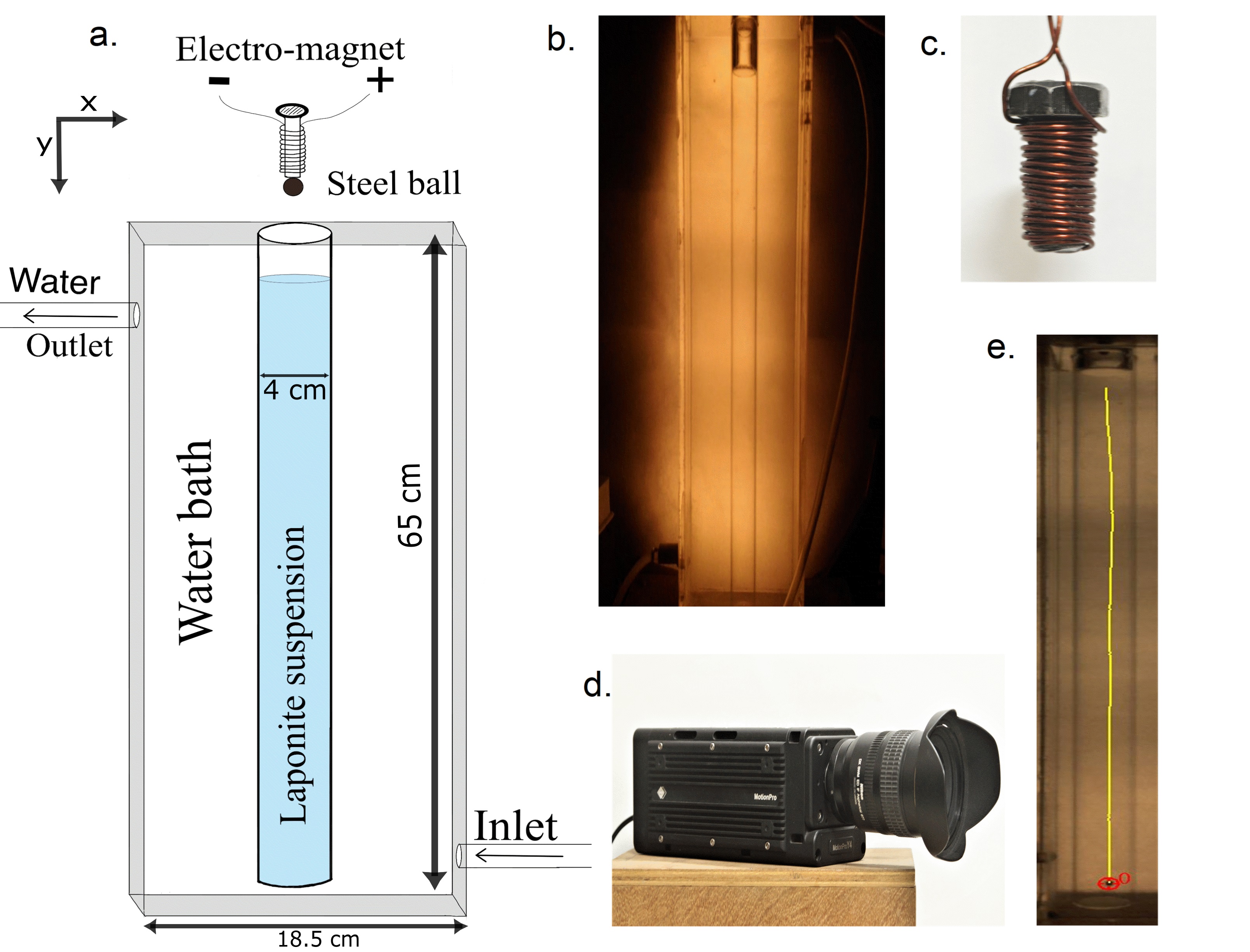}
\caption{ (a) Schematic diagram of the experimental set-up for ball drop experiments. (b) The cylindrical tube (length $65$ cm and diameter $4$ cm) filled with Laponite suspension placed inside a water bath at $25^ \circ$C. (c) The electro-magnet used to drop steel balls. (d) High speed camera - IDT Motion Pro Y4-S2. (e) Trajectory of a falling ball in the viscometer displayed in (b), tracked using Video Spot Tracker (CISMM), is shown by a yellow line.}
\end{figure}

\begin{figure}
\includegraphics[scale=.7]{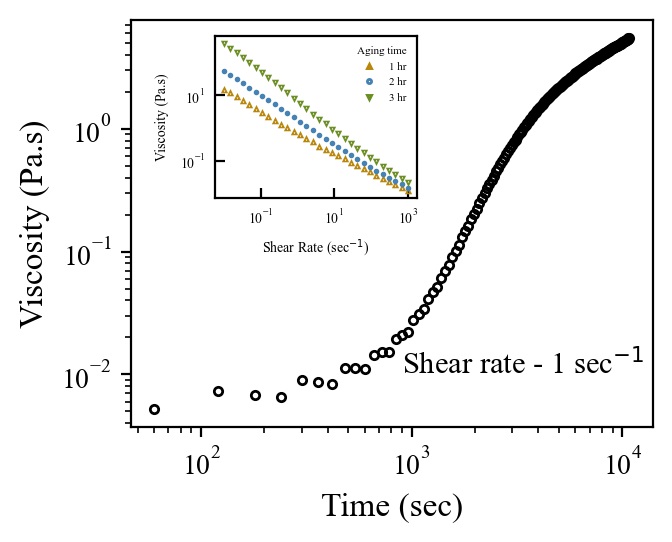}
\caption{ The time evolution of viscosity of a Laponite suspension (concentration 2.8\% $w/v$). In the inset, viscosity is plotted with shear rate for  2.8\% $w/v$ Laponite suspensions at different aging times.} 
\end{figure}

\subsection{Falling ball viscometry}
The schematic diagram of our falling ball viscometer is shown in Fig.1(a). For all the experiments reported here, the viscometer was inserted in a water bath whose temperature is maintained at 25$^\circ $C by a temperature controller (Polyscience Digital Inc.). After filtration, a freshly prepared Laponite suspension (approximately 600 ml) was loaded into the cylindrical viscometer shown in Fig.1(b). Once sample loading was complete, the top of the cylindrical tube was covered with parafilm to isolate the Laponite suspension from the atmosphere. The suspension was left to age, and its micro-structure was allowed to develop spontaneously due to the physical aging process. The aging time of the Laponite suspension is assigned the value $t_w$ = 0 as soon as sample loading was completed. Aging time is measured from the moment the viscometer was loaded till the time the ball was dropped in the viscometer. The time taken for the ball to fall through the suspension (less than 1 second)  is negligible compared to the aging times of the Laponite suspensions (1-3 hours) used in this study. The sample age was monitored continuously using a stopwatch for the entire duration of the experiment. The parafilm that was used to cover the Laponite suspension immediately after loading was completed was removed just before the ball was dropped. An electro-magnet shown in Fig.1(c) was used to drop steel balls along the vertical axis of the tube and was positioned at a fixed height of 1.5 cm above the surface of the Laponite suspension. This method of ball-drop was adopted since Laponite suspensions, particularly of higher ages, were seen to stick to the surface of the ball when the latter was immersed entirely in the suspension. The instant at which the falling ball enters the Laponite medium is recorded as time $t=0$. The corresponding ball velocity is designated as initial velocity, $V_0$, which is a fitting parameter in our model. Experiments were performed at several aging times corresponding to different structural states of the Laponite suspensions. After time $t_w$, the parafilm cover was removed from the top and steel balls (density $7815 \pm50$ kg/m$^3$) of different sizes were dropped using the electro-magnet while minimizing any rotation. The path followed by the ball was recorded with an IDT Motion Pro Y4-S2 high speed camera (Fig.1(d)) at 100 frames per second for all balls. The position of the falling ball was tracked from the images recorded by the camera with Video Spot Tracker (CISMM). Laponite suspensions were filled in the falling ball viscometer upto heights of 56-57 cm. According to Tanner\cite{Tanner}, end effects are insignificant if the ball is away from the bottom of the container by a distance that is more than the latter's radius. Following this, and since the radius of our viscometer is 2 cm, we have tracked the falling balls for a maximum of 54 cm from the top of the container (or 2 cm from the bottom). The yellow line in Fig.1(e) is a representative ball trajectory and was constructed from the position of the ball centroid in each frame. Horizontal displacement of the ball is estimated in Supplementary Materials (Fig.S2) and the horizontal velocity is seen to be negligible compared to the vertical velocity.The velocity of the falling ball was calculated from the position $vs.$ time data by time-differentiating the positions of the ball centroids using a central differential formula \cite{Garcia}. The error in velocity was estimated using the formula \cite{Feng2011}, $\delta V = \sqrt{\delta r_j^2 + \delta r_{j+1}^2} /\Delta t $, where $j$ is the frame number and the uncertainty in position ($\delta r =  \sqrt{\delta x^2 + \delta y^2}$) is the fitting error in estimating the mean of the intensity profile of the ball along the $y$ and  $x$ directions (direction of the fall of the ball and the direction perpendicular to it, i.e., parallel to the plane of the camera, respectively) using the Gaussian function. The time interval, $\Delta t$, was calculated from the frame rate of the camera. The data was fitted to a mathematical model using Jupyter notebook (Python v3.7). Experiments were performed using Newtonian fluids of viscosities comparable to those of the Laponite suspensions used here (Fig.S3 and Table.TS1 in Supplementary Material). The effects of the viscometer wall\cite{Kehlenbeck, Chhabra} and Reynolds number on the motion of the falling ball have been estimated from these experiments (Fig.S4 and Table.TS2 in Supplementary Material). The effects of the wall were found to be negligible in these measurements. By incorporating the Schiller Naumann correction \cite{CoulsonRichardson, goossens2019}, we found a good match between viscosity values obtained in falling ball and rheometric measurements.

\subsection{Rheology}
For rheological measurements \cite{Macosko, Ewoldt}, a stress controlled Anton Paar MCR-501 rheometer was used. The temperature was fixed at $25^\circ$C using a water circulation system. A double gap geometry (DG - $26.7$, outer diameter = $13.796$ mm, inner diameter = $11.915$ mm, cell height = $42$ mm, sample volume = $3.8$ ml) was used for our experiments. Aqueous suspensions of dried Laponite of concentration 2.8 \% $w/v$ were stirred for $1$ hr at $400$ rpm to break up big particle clusters. For each experimental run, the sample was loaded slowly into the annular region of the geometry using a syringe. Silicone oil of viscosity 5 cSt was used as solvent trap oil to minimize sample evaporation. A shear rate of 500 sec$^{-1}$ was applied for 5 minutes to break any remaining particle clusters.  This procedure ensures an identical initial state and is essential for the acquisition of reproducible data. A low shear (1 sec$^{-1}$) was applied for subsequent data acquisition and a freshly prepared sample was allowed to age spontaneously in the rheometer geometry upto a predetermined $t_w$ before the start of each measurement.

\section{Results}

\subsection{Time dependence of viscosity}

\begin{figure}
\includegraphics[scale=.295]{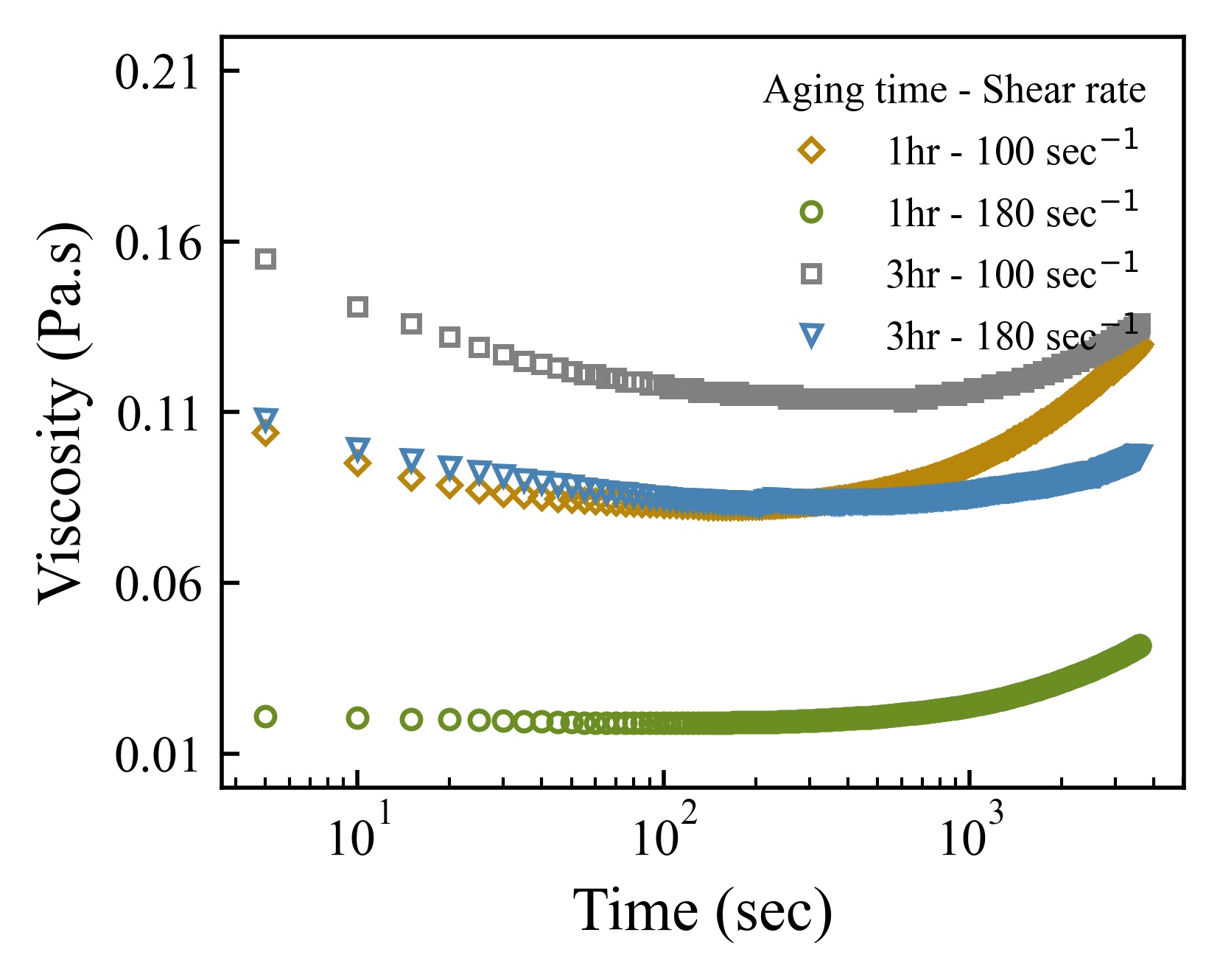}
\caption{Time dependent viscosity of 2.8\% $w/v$ Laponite suspensions, of aging times ($t_w$) $1$ hr and $3$ hr, subjected to constant shear rates of $100$ and $180$ sec$^{-1}$ at $25^\circ$C. } 
\end{figure}

\begin{figure}
\includegraphics[scale=.291]{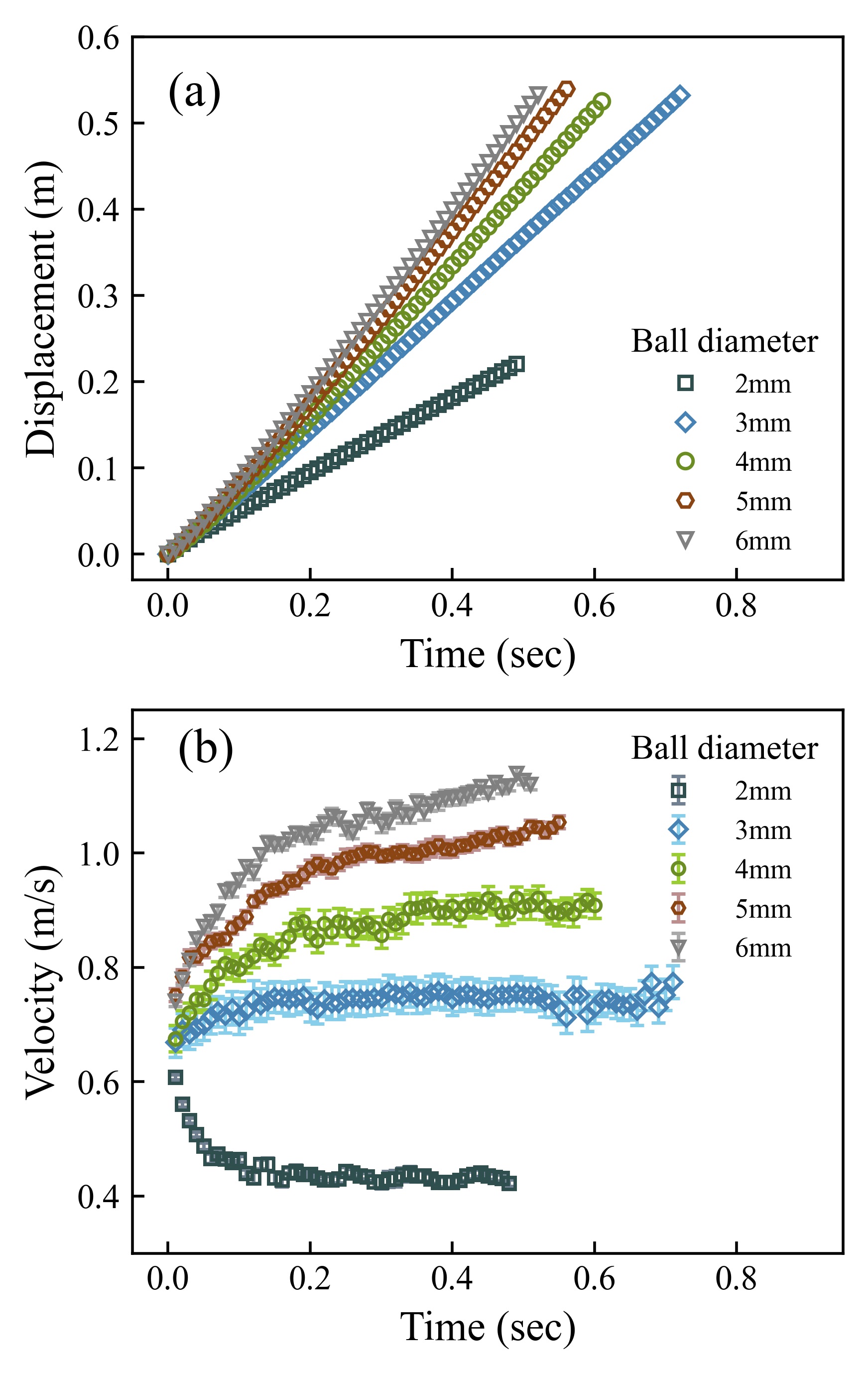}
\caption{(a) Displacement $vs.$ time of the steel balls (diameter = $2$, $3$, $4$, $5$, $6$ mm) falling through Laponite suspensions of concentration $2.8 \%$ $w/v$ at $1$ hr aging time ($t_w$). (b) Velocity $vs.$ time is calculated from the displacement data in (a). The error bars are calculated using the protocol specified for balls falling in a medium \cite{Feng2011}.}
\end{figure}

As discussed earlier, the time-dependent evolution of inter-particle screened repulsive forces in Laponite suspensions gives rise to a gradual and continuous micro-structural build-up. Consequently, Laponite suspensions show physical aging which manifests in experiments as a spontaneous evolution of the viscoelastic moduli. In Fig.2, the viscosity of a 2.8\% $w/v$ Laponite suspension, recorded with time under a low shear rate (1 sec$^{-1}$) for 3 hr, is displayed. The viscosity is seen to spontaneously increase with time, signalling the progress of the physical aging phenomenon. The viscosities of Laponite suspensions of aging times $t_w$ = 1,2,3 hours are plotted in the inset of Fig.2. As expected\cite{gueslin2009}, the viscosity decreases with shear rate for all sample aging times explored here. The shear stress vs. shear rate plots for Laponite suspensions at ages $t_w =$ 1,2,3 hrs are shown in Fig.S5 of Supplementary Material. Time-dependent changes in the viscosity of Laponite suspensions are measured for two different shear rates $100$ and $180$ sec$^{-1}$ and are plotted in Fig.3. It is seen from this figure that the imposition of constant shear rates results in a decrease in the suspension viscosity at earlier times. This observation can be attributed to the breakup of sample microstructures by the high shear rates that are imposed in these experiments. At longer times, the viscosity shows an increase due to the increased rate of structural recovery. These competing effects give rise to a non-monotonic viscosity profile that is sensitive to sample age and the imposed shear rate.
 
\begin{figure*}
\includegraphics[scale=.29]{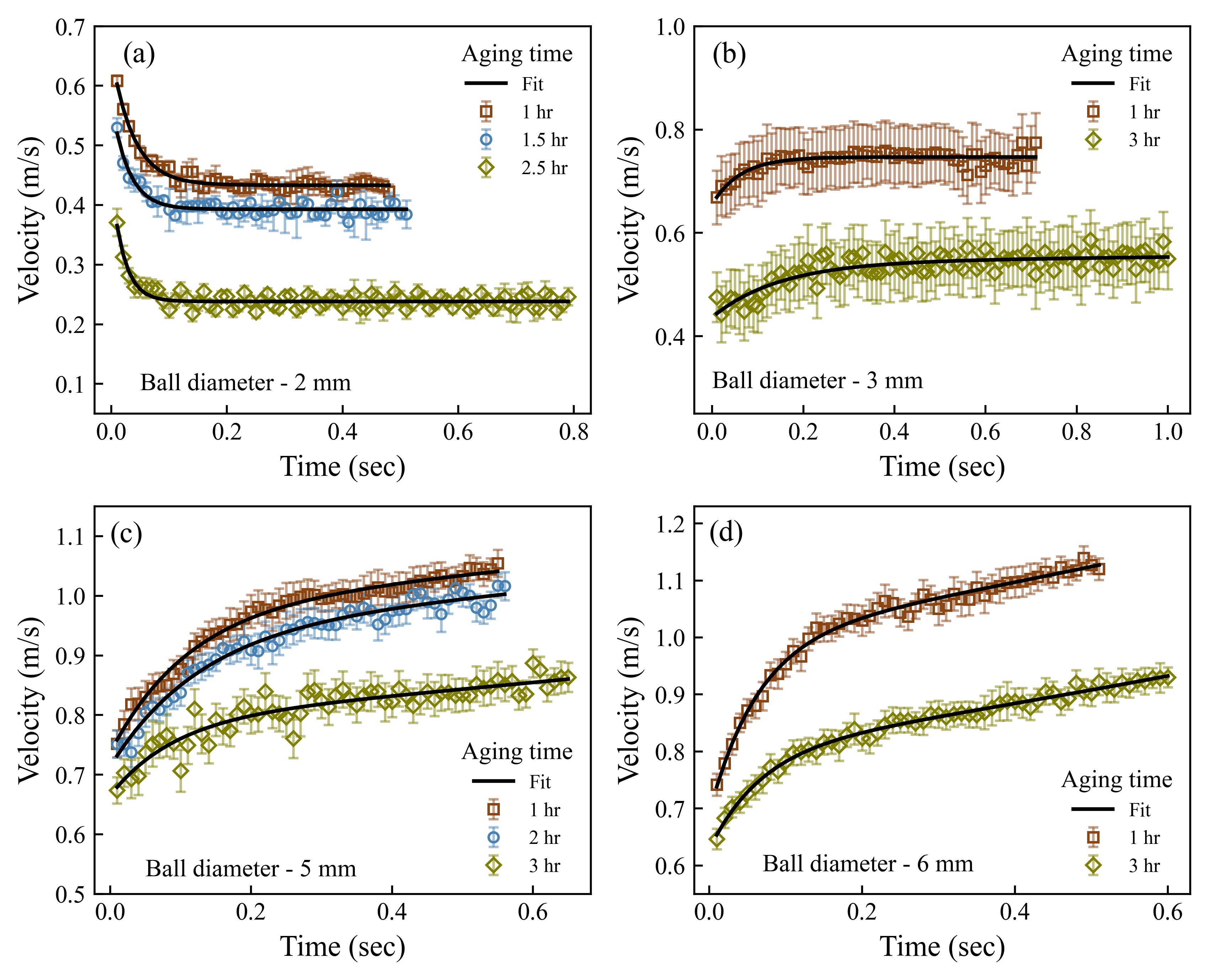}
\caption{In (a), (b), (c) and (d), velocities of the ball centroids for balls of diameters 2mm , 3mm , 5mm and 6mm respectively, falling through Laponite suspensions of different aging times ($t_w$), are plotted. Coloured symbols represent experimental data, coloured vertical bars are the corresponding error bars \cite{Feng2011} and continuous black lines are the fits to Eqn.(7).}
\end{figure*}

\subsection{Falling ball experiments}
The experiments reported here track millimeter-sized steel balls falling through spontaneously evolving Laponite suspensions. Freshly prepared Laponite suspensions that are allowed to age for $t_w$ =1,2,3 hrs are used. The viscosity of the suspension increases gradually due to the physical aging mechanism discussed earlier. A ball that is dropped through an aging viscoelastic suspension is expected to induce shear thinning of the medium due to break up of the fragile structures present in its path. For all the experiments reported here, Laponite suspension concentration is kept fixed at  $2.8 \%$ $w/v$. 

The extent of shear thinning of the surrounding medium due to passage of the ball can be evaluated approximately by relating it to the data plotted in Fig.3. The shear rate imposed by the ball is estimated using the relation $\dot{\gamma}= V/D$, where $V$ and $D$ are, respectively, the instantaneous velocity and the diameter of the falling ball. For the values of velocity, V, measured here and the diameters, D, of all the balls used, the shear rates imposed by the balls always lie within the range 100-200 sec$^{-1}$ in our experiments. A comparison with Fig.3 shows that the aging Laponite samples lie in the shear-thinning regime within this range of shear rates. A ball falling through a Laponite suspension would therefore disintegrate the local structures in its path.

Fig.4 shows representative data when balls are dropped in a Laponite suspension that has been aged for 1 hr. In Fig.4(a), the displacements of balls having different diameters (2mm, 3mm, 4mm, 5mm and 6mm ) are plotted. The overall slopes of the displacement curves tend to decrease with decreasing ball diameters. The velocities are plotted \textit{vs.} time for different ball sizes in Fig.4(b). As expected, the larger balls with higher inertia travel with greater velocities through the Laponite suspension medium. The ball of diameter 2mm, released at a high non-zero initial velocity, eventually reaches a lower terminal velocity. In contrast, the velocities of the balls of diameters 3-6 mm increase continuously with time throughout our experimental time window, with the increase becoming more prominent for the larger balls. Error in velocity due to tracking is calculated \cite{Feng2011} and plotted with vertical bars.

Data for all the ball-drop experiments, acquired for balls of sizes between 2 mm and 6 mm falling through Laponite suspensions of different ages, are plotted in Fig.5. We observe from Fig.5(a) that the velocity of the 2 mm ball decays with time and attains a steady state terminal velocity during the experimental time window for all Laponite suspensions at aging times, $t_w$= 1 hr, 1.5 hr, 2.5 hr. For all the Laponite aging times investigated here, the velocities of the balls of larger diameters (3mm, 5mm and 6mm) continue to increase during the experimental time window, with the increase being maximum for the largest ball (Figs.5(b-d)). Such continuous increase in velocity without attainment of terminal velocity in the experimental time window accessible in our experiments indicates the prevalence of medium destructuring over structural recovery under these conditions. As the sample age increases, the ball velocity decreases in all the experiments. Vertical bars with lighter shades of the symbol colors are plotted to represent the errors in Fig.5.

\subsection{Mathematical model}
Moore (1959) \cite{moore1959} proposed a mathematical model to explain the micro-structural evolution of thixotropic materials. In this model, the time-dependent structure of an aging fluid is represented by a dimensionless structural parameter, $\lambda$. This model incorporates $\lambda$ with the following physical interpretation: $\lambda = 0$ for a fully broken or destructured state of the fluid (or when the structure is yet to form at the microscopic level), while $\lambda = 1$ for a fully structured fluid state, \textit{i.e.}, when the structure is not broken at all. 

In an aging Laponite suspension that is locally destructured by a ball falling through it, the time evolution of the structural parameter, $\lambda$, has to be governed by restructuring/structural recovery and destructuring/structural breakdown of the suspension:

\begin{equation}
\frac{d\lambda}{dt} = F(\lambda)-G(\lambda,\dot{\gamma})  \label{two}
\end{equation}

In the above equation, $F(\lambda$) and $G(\lambda,\dot{\gamma})$ are associated with the rates of the restructuring and destructuring processes that progress simultaneously in a Laponite suspension having an instantaneous state $\lambda$. Eqn.2 can be simplified by considering a constant restructuring rate $F=1/\theta$ and a destructuring function $G= \alpha \lambda \dot{\gamma}$ that varies linearly with local shear rate\cite{ferrior2004, toorman1997}, i.e., $d\lambda/dt = 1/\theta-\alpha \lambda \dot{\gamma}$. Here, $\dot{\gamma}$ is the shear rate imposed by the falling ball and $1/\theta$ and $\alpha$ are parameters related to the restructuring and destructuring rates of the medium respectively.

The restructuring rate of the Laponite suspension is decided by its physical properties such as aging time, concentration and instantaneous state. As the Laponite suspension ages with time, the structure evolves identically throughout the medium. In the present scenario, since the structural length scales of interest are much larger than the size of a single Laponite particle, the viscosity can be assumed to be equal at all points in the undisturbed sample. The high shear rates of approximately 100-200 sec$^{-1}$ experienced by the regions in the immediate neighborhood of the falling ball are expected to result in shear thinning of the surrounding Laponite medium due to the breakdown of the fragile microscopic structures. This can be seen in Fig.3 where we note that structural recovery takes over only after 1000 seconds when suspensions are subjected to these high shear rates. The time of fall of the ball is less than 1 second and therefore smaller than the time scale over which the structural parameter is expected to change due to the physical aging process. The destructuring due to the passage of the ball is expected to dominate over the restructuring process and $F(\lambda)$ in Eqn.2 can be neglected. To model the destructuring of the medium in our falling ball experiments, we make the following assumptions:

i) The ball falls inside the sample while shearing only the immediate vicinity of its surface. During the passage of the ball, it interacts with each point in the medium that lies in its path. Therefore the time evolution of the structural parameter of the medium, $d\lambda_f/dt$, should be proportional to the time taken by the ball to travel its diameter, $D/V$.

ii) As higher ball velocity indicates increased disruption in the structure of the medium surrounding the ball, the time evolution of the structural parameter is assumed proportional to the instantaneous ball velocity, $V$, during the fall.
\\

We can therefore simplify Eqn.2 and propose the following modified kinetic equation for the time-dependent structural evolution in Laponite suspensions,
\begin{equation}
\frac{d\lambda_f}{dt} \propto - V \frac{D}{V}  \label{four}
\end{equation}

\begin{table*}
	\begin{center}
    \caption{Parameters measured by fitting Eqn.7 to the data in Fig.5.}
    \label{tab:table2}
    \begin{tabularx}{1.0\textwidth}{|>{\centering\arraybackslash}X|>{\centering\arraybackslash}X|>{\centering\arraybackslash}X|>{\centering\arraybackslash}X|>{\centering\arraybackslash}X|>{\centering\arraybackslash}X|} 
    \hline
      Ball diameter (mm) & Aging time (hr)& Initial velocity $V_0$ $(m/s)$ & $W$ $(m/s)$ & $1/\tau$ $(s^{-1})$ & Destructuring rate $\beta$ $(s^{-1})$\\
      \hline
      2 & 1   & 0.653 $\pm$ 0.010 & 0.433 $\pm$ 0.004 & 26.077 $\pm$ 2.255 & 1.07e-17 $\pm$ 0.031\\
      2 & 1.5 & 0.570 $\pm$ 0.019 & 0.392 $\pm$ 0.005 & 33.232 $\pm$ 5.245 & 1.64e-19 $\pm$ 0.037\\
      2 & 2.5 & 0.439 $\pm$ 0.024 & 0.238 $\pm$ 0.003 & 45.801 $\pm$ 6.843 & 7.79e-16 $\pm$ 0.026\\
      3 & 1   & 0.655 $\pm$ 0.011 & 0.747 $\pm$ 0.006 & 16.033 $\pm$ 3.814 & 1.11e-15 $\pm$ 0.011\\
      3 & 3   & 0.436 $\pm$ 0.010 & 0.543 $\pm$ 0.013 & 6.819  $\pm$ 2.001 & 0.017 $\pm$ 0.030\\
      4 & 1   & 0.661 $\pm$ 0.009 & 0.876 $\pm$ 0.013 & 10.882 $\pm$ 1.499 & 0.066 $\pm$ 0.029\\
      5 & 1   & 0.736 $\pm$ 0.007 & 0.982 $\pm$ 0.016 & 9.374  $\pm$ 1.130 & 0.105 $\pm$ 0.031\\
      5 & 2   & 0.713 $\pm$ 0.009 & 0.947 $\pm$ 0.029 & 8.208  $\pm$ 1.719 & 0.103 $\pm$ 0.053\\
      5 & 3   & 0.664 $\pm$ 0.014 & 0.793 $\pm$ 0.015 & 11.757 $\pm$  3.456 & 0.119 $\pm$ 0.034\\
      6 & 1   & 0.691 $\pm$ 0.010 & 1.001 $\pm$ 0.008 & 16.261 $\pm$ 1.206 & 0.218 $\pm$ 0.018 \\
      6 & 3   & 0.627 $\pm$ 0.006 & 0.802 $\pm$ 0.004 & 15.872 $\pm$ 1.242 & 0.232 $\pm$ 0.010 \\
      \hline
    \end{tabularx}
    \end{center}
\end{table*} 

In Eqn.3, the time dependent structural parameter  of the aging Laponite suspension for falling ball experiments is denoted by $\lambda_f$. The negative sign is incorporated to account for shear-thinning due to disruption of the fragile suspension structures by the falling ball. Eqn.3 reduces to:

\begin{align}
\frac{d\lambda_f}{dt} & = - \epsilon D \\ \
\implies \lambda_f & = \lambda_0(1 - \beta t)   \label{five}
\end{align}
where $\epsilon$ is a proportionality constant of dimension (length $\times$ time)$^{-1}$, $\lambda_f = \lambda_0$ at time $t$ = 0 and the rate of destructuring is $\beta = \epsilon D/\lambda_0$. The viscoelastic, structured nature of the suspension can be incorporated in the expression for terminal velocity, $V_{ter}$ obtained from Eqn.1, by considering the functional form\cite{gumulya2014} for viscosity $\eta=\eta_0\lambda_f^n $ where $\eta_0$ is the zero-shear viscosity  and $n=1$ for a Laponite suspension \cite{ferrior2004}. Viscosity of the suspension at each point changes with its structural parameter. The expression for the structure dependent terminal velocity, $V_{\lambda}$, of the ball falling through a thixotropic Laponite suspension can therefore be written as:
\begin{align}
V_\lambda(t) & = \frac{(\rho_s - \rho)g D^2}{18\eta(t)} \nonumber \\
   & = \frac{(\rho_s - \rho)g D^2}{18\eta_0 \lambda_0(1-\beta t)}
\end{align}
Here, we have used Eqn.5 and the expression for viscosity of Laponite suspensions \cite{gumulya2014} to obtain the expression for instantaneous viscosity, $\eta(t) = \eta_0 \lambda_0(1-\beta t)$. Using Eqn.(1) and (6), we rewrite the equation of motion of a ball falling through a thixotropic suspension with an initial velocity $V_0$:
\begin{align}
V & = V_\lambda + (V_0 - V_\lambda)\exp(-t/\tau) \nonumber \\
& = \frac{W}{(1 - \beta t)}+ (V_0 - \frac{W}{(1 -\beta t) })\exp(-t/\tau)  \label{seven}
\end{align}
where $W =(\rho_s - \rho)g D^2/18\eta_0 \lambda_0$, $\beta$, $V_0$, $\tau$ are the fitting parameters. Here, $\tau = m/3\pi\lambda_0\eta_0 D$ where $m$ = mass of the ball, $\eta_0$ = zero-shear viscosity of the sample, $D$ = diameter of the ball.  

The constant breakdown of the micro-structure as the ball falls through the suspension leads to shear thinning at a rate parametrized by $\beta$ in Eqn.5. Naturally, for a Newtonian sample, we expect the value of $\beta$ to be zero. The data for balls of different sizes falling through Laponite suspensions of different aging times (Figs.5(a-d)) are fitted to Eqn.7 (solid lines in Fig.5) and the destructuring rate, $\beta$, is obtained as one of the fitting parameters. All the fitting parameters are tabulated in Table I. In Fig.6, $\beta$ is plotted for ball sizes between 2 mm - 6 mm for different aging times. Only the fitting errors are considered in this plot. It is clear from the figure that $\beta$ decreases with ball size and is approximately zero for the smallest ball of diameter 2 mm studied here. When larger balls are dropped through Laponite suspensions, the destructuring rates increase with ball size and stay nearly unchanged over the entire range of sample ages. The approximate insensitivity of $\beta$ to changes in sample age $t_w$ arises from the predominance of destructuring processes over structural recovery in the freshly prepared Laponite suspensions used here. Our assumption that shear thinning of the medium dominates over structural recovery in our falling ball experiments is therefore adequate to accurately compute the ball trajectories. Furthermore, Eqn.7 explains our experimental data irrespective of the instantaneous state of the suspension. 

\begin{figure}
\includegraphics[scale=.29]{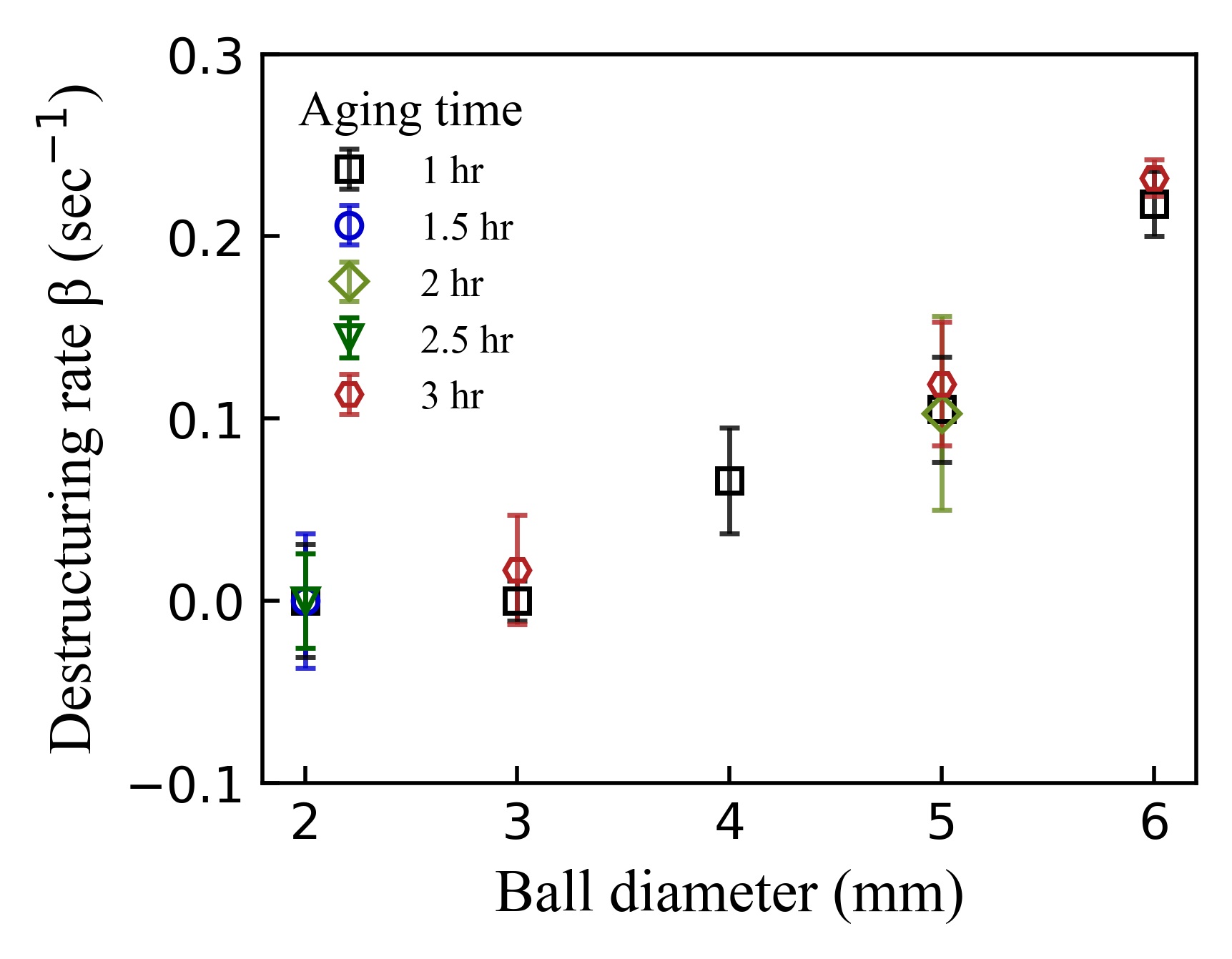}
\caption{Destructuring rate, $\beta$, $vs.$ ball diameter, $D$, is plotted for different aging times ($t_w$) of Laponite suspensions.}
\end{figure}

\section{Conclusions}
The destructuring of a Laponite suspension due to a ball falling through it is studied in this work. We show that the motion of solid steel balls through spontaneously evolving colloidal suspensions of Laponite is very different from the motion of a solid object in a Newtonian fluid. 

Falling ball experiments are performed by dropping millimeter-sized balls in Laponite suspensions of different ages. In contrast to a ball falling through a Newtonian fluid, we find that the larger balls do not attain a steady state while falling through Laponite suspensions during the experimental time window. By modifying a mathematical model proposed by Moore \cite{moore1959} to account for balls falling over time scales much faster than the suspension aging times, we successfully extract the destructuring rates of Laponite suspensions of different ages for different ball sizes. Interestingly, the gradual increase of velocity within the experimental time window accessible in these measurements can also be explained using this model. Eventually at later times and for longer viscometer heights, the ball would have certainly reached a steady state. The restructuring contribution in Eqn.2 that we have neglected for the present measurements would then be non-negligible. However, trying to observe the motion of the ball till it reaches a steady state would require us to perform experiments with much longer cylinders and enormous sample volumes. We have therefore restricted our study by exploring the motion of the falling ball in a regime that can be described by the simplest mathematical model that quantifies the gradual increase of ball velocity due to shear thinning of the Lapaonite suspensions. Further, the initial velocity of the ball when it enters the medium is non-zero as the ball is released from a fixed height above the fluid surface. Further experiments need to be performed to determine the effect of initial velocity on the destructuring rate of the suspension.  

Our results are fairly general and are applicable in the study of the settling dynamics of macroscopic objects in a wide variety of thixotropic suspensions. Balls of different materials can be used to explore microscopic viscosities and destructuring rates of complex fluids over a large range of shear rates. We note here that while the viscosity measured by a rheometer reflects the response of the entire sample, falling ball viscometry explores the viscosity of the medium only in the neighborhood of the falling ball. We have parametrized the destructuring rate of the Laponite suspension medium in terms of a parameter $\beta$. Rheometry and falling ball viscometry probe the system differently, involve very different shear profiles, and are complementary to each other. A quantitative relation between $\beta$ and bulk viscosity of a non-Newtonian fluid obtained in rheological measurements is therefore well outside the scope of this paper.

 To conclude, falling ball viscometry, which involves the macroscopic measurement of the velocity of a ball falling through a medium, can be employed to successfully estimate the microscopic dynamics of fragile structures in shear thinning suspensions. The experiments reported in this paper provide important insights into the physical aging process in the presence of externally-imposed forces and contribute to our understanding of the motion of macroscopic objects through non-Newtonian fluids.
\\
\\

\noindent \textbf{Supplementary Material} - See supplementary material for velocity $vs.$ time data for millimetre sized balls falling through Newtonian glycerol-water mixtures and non-Newtonian Laponite suspensions of comparable viscosities, estimation of radial velocities of the falling balls, estimations of wall effects and the effects of the intermediate values of Reynolds numbers for balls dropped in glycerol-water mixtures of comparable viscosities, and the flow curve of 2.8\% $w/v$ Laponite suspensions of ages 1, 2 and 3 hrs.

\begin{acknowledgments}
We would like to thank Abhishek Dhar and Rama Govindarajan for useful discussions. We acknowledge CISMM at UNC-CH, supported by the NIH NIBIB (NIH 5-P41-RR02170) for the tracking software. We also acknowledge Sanjay Kumar Behera and Yash Rana for their help during the initial stages of the experiment.
\end{acknowledgments}
\noindent \textbf{Availability of data} - The data that support the findings of this study are available from the corresponding author upon reasonable request.

This article has been submitted to Physics of Fluids.

\bibliography{aipsamp}

\begin{thebibliography}{5}

\bibitem{stokes1851}G.G. Stokes (1851). On the effect of the internal friction of fluids on the motion of pendulums. Trans Cambridge philos. Soc. 9,8.

\bibitem{CoulsonRichardson}Harker, J. h.,  Richardson, J. F. and Backhurst, J.R. (2002). Chapter 3,  Coulson And Richardsons Chemical Engineering, Volume 2, 5Th Edition: Particle Technology And Separation Processes. Oxford: Butterworth-Heinemann.

\bibitem{brown2003}Brown, P. P., and Lawler, D. F. (2003). Sphere Drag and Settling Velocity Revisited. Journal of Environmental Engineering, 129(3), 222–231. https://doi.org/10.1061/(asce)0733-9372(2003)129:3(222).

\bibitem{goossens2019}Goossens, W. R. A. (2019). Review of the empirical correlations for the drag coefficient of rigid spheres. Powder Technology, 352, 350–359. https://doi.org/10.1016/j.powtec.2019.04.075.

\bibitem{khan1987}Khan, A. R., and Richardson, J. F. (1987). The resistance to motion of a solid sphere in a fluid. Chemical Engineering Communications, 62(1–6), 135–150. https://doi.org/10.1080/00986448708912056

\bibitem{little1976}Hart, F. X., and Little, C. A. (1976). Student investigation of models for the drag force. American Journal of Physics, 44(9), 872–878. https://doi.org/10.1119/1.10286

\bibitem{gueslin2006}Gueslin, B., Talini, L., Herzhaft, B., Peysson, Y., and Allain, C. (2006). Flow induced by a sphere settling in an aging yield-stress fluid. Physics of Fluids, 18(10), 103101. https://doi.org/10.1063/1.2358090

\bibitem{Freund} Freund, J. B., Kim, J., and Ewoldt, R. H. (2018). Field sensitivity of flow predictions to rheological parameters. Journal of Non-Newtonian Fluid Mechanics, 257, 71–82. https://doi.org/10.1016/j.jnnfm.2018.03.013

\bibitem{Guvendiren2012} Guvendiren, M., Lu, H. D., and Burdick, J. A. (2012). Shear-thinning hydrogels for biomedical applications. Soft Matter, 8(2), 260–272. https://doi.org/10.1039/c1sm06513k.

\bibitem{Bonn2010} Fall, A., Paredes, J., and Bonn, D. (2010). Yielding and Shear Banding in Soft Glassy Materials. Physical Review Letters, 105(22). https://doi.org/10.1103/physrevlett.105.225502

\bibitem{Fong1996}Fong, C. F. C. M., Turcotte, G., and De Kee, D. (1996). Modelling steady and transient rheological properties. Journal of Food Engineering, 27(1), 63–70. https://doi.org/10.1016/0260-8774(94)00077-m

\bibitem{Giannokostas} Giannokostas, K., Moschopoulos, P., Varchanis, S., Dimakopoulos, Y., and Tsamopoulos, J. (2020). Advanced constitutive modeling of the thixotropic elasto-visco-plastic behavior of blood: Description of the model and rheological predictions. Materials, 13(18). https://doi.org/10.3390/ma13184184

\bibitem{Clark} Clark, C. C., Aleman, J., Mutkus, L., and Skardal, A. (2019). A mechanically robust thixotropic collagen and hyaluronic acid bioink supplemented with gelatin nanoparticles. Bioprinting, 16. https://doi.org/10.1016/j.bprint.2019.e00058

\bibitem{Liu} Liu, Y., Ling, S., Wang, S., Chen, X., and Shao, Z. (2014). Thixotropic silk nanofibril-based hydrogel with extracellular matrix-like structure. Biomaterials Science, 2(10), 1338–1342. https://doi.org/10.1039/c4bm00214h

\bibitem{Chhabra}Chhabra, R. P. (2006). Bubbles, Drops, and Particles in Non-Newtonian Fluids (Chemical Industries) (2nd ed.). CRC Press.

\bibitem{Zhuang2019} Zhuang, G., Zhang, Z., and Jaber, M. (2019). Organoclays used as colloidal and rheological additives in oil-based drilling fluids: An overview. Applied Clay Science, 177, 63–81. https://doi.org/10.1016/j.clay.2019.05.006

\bibitem{Mrokowska2019} Mrokowska, M. M., and Krztoń-Maziopa, A. (2019). Viscoelastic and shear-thinning effects of aqueous exopolymer solution on disk and sphere settling. Scientific Reports, 9(1). https://doi.org/10.1038/s41598-019-44233-z

\bibitem{jayaraman2003}Jayaraman, A., and Belmonte, A. (2003). Oscillations of a solid sphere falling through a wormlike micellar fluid. Physical Review E, 67(6). https://doi.org/10.1103/physreve.67.065301

\bibitem{kumar2012}Kumar, N., Majumdar, S., Sood, A., Govindarajan, R., Ramaswamy, S., and Sood, A. K. (2012). Oscillatory settling in wormlike-micelle solutions: bursts and a long time scale. Soft Matter, 8(16), 4310. https://doi.org/10.1039/c2sm25077b

\bibitem{Handzy2013}Handzy, N. Z., and Belmonte, A. (2004). Oscillatory Rise of Bubbles in Wormlike Micellar Fluids with Different Microstructures. Physical Review Letters, 92(12). https://doi.org/10.1103/physrevlett.92.124501

\bibitem{Briscoe1992}Briscoe, B. J., Glaese, M., Luckham, P. F., and Ren, S. (1992). The falling of spheres through Bingham fluids. Colloids and Surfaces, 65(1), 69–75. https://doi.org/10.1016/0166-6622(92)80176-3

\bibitem{Kroon1998}Kroon, M., Vos, W., and Wegdam, G. (1998). Structure and formation of a gel of colloidal disks. Physical Review E, 57(2), 1962–1970. https://doi.org/10.1103/physreve.57.1962

\bibitem{Ali2013} Ali, S., and Bandyopadhyay, R. (2013). Use of Ultrasound Attenuation Spectroscopy to Determine the Size Distribution of Clay Tactoids in Aqueous Suspensions. Langmuir, 29(41), 12663–12669. https://doi.org/10.1021/la402478h

\bibitem{Tawari2001} Tawari, S. L., Koch, D. L., and Cohen, C. (2001). Electrical Double-Layer Effects on the Brownian Diffusivity and Aggregation Rate of Laponite Clay Particles. Journal of Colloid and Interface Science, 240(1), 54–66. https://doi.org/10.1006/jcis.2001.7646

\bibitem{Joshi2007} Joshi, Y. M. (2007). Model for cage formation in colloidal suspension of laponite. The Journal of Chemical Physics, 127(8), 081102. https://doi.org/10.1063/1.2779026

\bibitem{Bonn} Bonn, D., Tanase, S., Abou, B., Tanaka, H., and Meunier, J. (2002). Laponite: Aging and shear rejuvenation of a colloidal glass. Physical Review Letters, 89(1), 157011–157014. https://doi.org/10.1103/PhysRevLett.89.015701

\bibitem{Saha2015} Saha, D., Bandyopadhyay, R., and Joshi, Y. M. (2015). Dynamic Light Scattering Study and DLVO Analysis of Physicochemical Interactions in Colloidal Suspensions of Charged Disks. Langmuir, 31(10), 3012–3020. https://doi.org/10.1021/acs.langmuir.5b00291

\bibitem{Jatav2014} Jatav, S., and Joshi, Y. M. (2014). Chemical stability of Laponite in aqueous media. Applied Clay Science, 97–98, 72–77. https://doi.org/10.1016/j.clay.2014.06.004

\bibitem{ruzika2011}Ruzicka, B., and Zaccarelli, E. (2011). A fresh look at the Laponite phase diagram. Soft Matter, 7(4), 1268. https://doi.org/10.1039/c0sm00590h

\bibitem{Cummins2007} Cummins, H. Z. (2007). Liquid, glass, gel: The phases of colloidal Laponite. Journal of Non-Crystalline Solids, 353(41–43), 3891–3905. https://doi.org/10.1016/j.jnoncrysol.2007.02.066

\bibitem{Shahin2010} Shahin, A., and Joshi, Y. M. (2010). Irreversible Aging Dynamics and Generic Phase Behavior of Aqueous Suspensions of Laponite. Langmuir, 26(6), 4219–4225. https://doi.org/10.1021/la9032749

\bibitem{Angelini2013}Angelini, R., Zulian, L., Fluerasu, A., Madsen, A., Ruocco, G., and Ruzicka, B. (2013). Dichotomic aging behaviour in a colloidal glass. Soft Matter, 9(46), 10955. https://doi.org/10.1039/c3sm52173g

\bibitem{tabuteau2007}Tabuteau, H., Oppong, F. K., Bruyn, J. R., and Coussot, P. (2007). Drag on a sphere moving through an aging system. Europhysics Letters (EPL), 78(6), 68007. https://doi.org/10.1209/0295-5075/78/68007

\bibitem{Fazilati2017}Fazilati, M., Maleki-Jirsaraei, N., Rouhani, S., and Bonn, D. (2017). Quasi-periodic and irregular motion of a solid sphere falling through a thixotropic yield-stress fluid. Applied Physics Express, 10(11), 117301. https://doi.org/10.7567/apex.10.117301

\bibitem{moller2006}Møller, P. C. F., Mewis, J., and Bonn, D. (2006). Yield stress and thixotropy: on the difficulty of measuring yield stresses in practice. Soft Matter, 2(4), 274. https://doi.org/10.1039/b517840a

\bibitem{Putz2008} Putz, A. M. V., Burghelea, T. I., Frigaard, I. A. and Martinez, D. M. \textsl{Settling of an isolated spherical particle in a yield stress shear thinning fluid.} Physics of Fluids 20, (2008).

\bibitem{moore1959}Moore F. (1959). The rheology of ceramic slips and bodies. Trans Brit Ceramic Soc 58:470-494.

\bibitem{Tanner} Tanner, R. I. (1963). End effects in falling-ball viscometry. Journal of Fluid Mechanics, 17(2), 161–170. https://doi.org/10.1017/S002211206300121X

\bibitem{Garcia} A.L. Garcia \textsl{Numerical Methods for Physics, second ed.} Prentice-Hall Upper Saddle River, NJ, 2000.

\bibitem{Feng2011}Feng, Y., Goree, J., and Liu, B. (2011). Errors in particle tracking velocimetry with high-speed cameras. Review of Scientific Instruments, 82(5), 053707. https://doi.org/10.1063/1.3589267

\bibitem{Kehlenbeck} Kehlenbeck, R., and Di Felice, R. (1999).Empirical relationships for the terminal settling velocity of spheres in cylindrical columns. Chemical Engineering and Technology, 22(4), 303–308. https://doi.org/10.1002/(SICI)1521-4125(199904)22:4<303::AID-CEAT303>3.0.CO;2-8

\bibitem{Chhabra} Chhabra, R. P., Agarwal, S., and Chaudhary, K. (2003). A note on wall effect on the terminal falling velocity of a sphere in quiescent Newtonian media in cylindrical tubes. Powder Technology, 129(1–3), 53–58. https://doi.org/10.1016/S0032-5910(02)00164-X

\bibitem{Macosko} Macosko, C. W. (1996). Rheology: Principles, Measurements and Applications. Powder Technology (Vol. 86, p. 313).

\bibitem{Ewoldt} Ewoldt, R. H., Johnston, M. T., and Caretta, L. M. (2015). Experimental Challenges of Shear Rheology: How to Avoid Bad Data (pp. 207–241). https://doi.org/10.1007/978-1-4939-2065-5 6.

\bibitem{gueslin2009}Gueslin, B., Talini, L., and Peysson, Y. (2009). Sphere settling in an aging yield stress fluid: link between the induced flows and the rheological behavior. Rheologica Acta, 48(9), 961–970. https://doi.org/10.1007/s00397-009-0376-6

\bibitem{ferrior2004}Ferroir, T., Huynh, H. T., Chateau, X., and Coussot, P. (2004). Motion of a solid object through a pasty (thixotropic) fluid. Physics of Fluids, 16(3), 594–601. https://doi.org/10.1063/1.1640372

\bibitem{toorman1997}Toorman, E. A. (1997). Modelling the thixotropic behaviour of dense cohesive sediment suspensions. Rheologica Acta, 36(1), 56–65. https://doi.org/10.1007/bf00366724

\bibitem{gumulya2014}Gumulya, M. M., Horsley, R. R., and Pareek, V. (2014). Numerical simulation of the settling behaviour of particles in thixotropic fluids. Physics of Fluids, 26(2), 023102. https://doi.org/10.1063/1.4866320

\end{thebibliography}

\appendix
\renewcommand\thefigure{\thesection.\arabic{figure}}
\setcounter{figure}{0}

\end{document}